\documentclass[aps,prl,twocolumn,amsmath,amssymb,nofootinbib,showpacs,superscriptaddress]{revtex4-1}
\usepackage[english]{babel}
\usepackage{latexsym}
\usepackage{graphics}
\usepackage{epsfig}
\usepackage{color}
\usepackage{bm}
\usepackage{amsmath}
\usepackage{amssymb}
\usepackage{tikz}  \usetikzlibrary{patterns}
\usepackage{braket}
\usepackage{comment}

\usepackage{lipsum} % just for the examples

\newcommand{\be}{\beta}
\newcommand{\ze}{\zeta}

\newcommand{\si}{\sigma}
\newcommand{\ep}{\epsilon}
\newcommand{\de}{\delta}
\newcommand{\al}{\alpha}

\renewcommand{\de}{\delta}
\newcommand{\De}{\Delta}

\newcommand{\pder}[2]{\dfrac{\partial#1}{\partial#2}}

\newcommand{\eps}{\epsilon_*}

\newcommand{\beq}{\begin{equation}}
\newcommand{\eeq}{\end{equation}}
\newcommand{\alig}[1]{\begin{align}#1\end{align}}
\newcommand{\aligns}[1]{\begin{align*}#1\end{align*}}

\newcommand\numberthis{\addtocounter{equation}{1}\tag{\theequation}}
\newcommand{\fr}[2]{\frac{#1}{#2}}

\newcommand{\tbf}[1]{\textbf{#1}}

\begin{document}

\title{Many-body localization in continuum systems: two-dimensional bosons}

\author{G. Bertoli}
\affiliation{LPTMS, CNRS, Univ. Paris-Sud, Universit\'e Paris-Saclay, Orsay 91405, France}

\author{B.L. Altshuler}
\affiliation{Physics Department, Columbia University, 538 West 120th Street, New York, New York 10027, USA}

\author{G.V. Shlyapnikov}
\affiliation{LPTMS, CNRS, Univ. Paris-Sud, Universit\'e Paris-Saclay, Orsay 91405, France}
\affiliation{SPEC, CEA, CNRS, Universit\'e Paris-Saclay, CEA Saclay, Gif sur Yvette 91191, France}
\affiliation{Russian Quantum Center, Skolkovo, Moscow Region 143025, Russia}
\affiliation{\mbox{Van der Waals-Zeeman Institute, University of Amsterdam, Science Park 904, 1098 XH Amsterdam, The Netherlands}}
%\affiliation{State Key Laboratory of Magnetic Resonance and Atomic and Molecular Physics,
%Wuhan Institute of Physics and Mathematics, Chinese Academy of Sciences, Wuhan 430071, China}

\date{\today}
\begin{abstract}
We demonstrate that many-body localization of two-dimensional weakly interacting bosons in disorder remains stable in the thermodynamic limit at sufficiently low temperatures. Highly energetic particles destroy the localized state only above a critical temperature, which increases with the strength of the disorder.  If the particle distribution is truncated at high energies, as it does for cold atom systems, the localization can be stable at any temperature.
\end{abstract}
\maketitle

After several decades since its original formulation, Anderson localization of quantum particles in disorder \cite{Anderson} continues to attract a lot of interest \cite{50years}. One of the key questions is the effect of interactions on the localization \cite{Fleishman,Altshuler}. It was established that localization may persist when interactions are present \cite{Basko, Gornyi1}, and the study of this phenomenon, known as many-body localization (MBL), is rapidly progressing \cite{ReviewHuse, ReviewAbanin}. A substantial amount of work, with a major role played by numerics, has been performed on lattice systems. The existence of the MBL phase has been proven in a rather general case of one-dimensional (1D) spin chains \cite{Imbrie}, and a growing number of experimental observations provided further relevant evidence \cite{Tanzi,Schreiber, Smith, Bordia, Choi, Bordia2, Rubio}. 

In continuum systems, there is no bound on the bandwidth as in lattice systems. The seminal work on the problem of MBL in continuum systems demonstrated that interacting particles can undergo many-body localization-delocalization transition (MBLDT), i.e. the transition from insulator to fluid state, and used an energy-independent single-particle localization length \cite{Basko,Gornyi1}. This condition was relaxed in a number of subsequent works, which took into account the growth of the localization length with energy \cite{Aleiner, Michal, Michal2, Nandkishore, Gornyi, Bertoli}. The question is whether the growth of the localization length eventually destabilizes the MBL phase, or whether the contribution of highly energetic states is hindered by the decrease in the thermal distribution function. 

In the recent work \cite{Bertoli}, it was shown that two-dimensional (2D) disordered bosons may display a finite-temperature insulator due to truncation of the energy distribution function at high energies, a generic phenomenon in cold atom systems \cite{walraven,ketterle}. The truncation ensures the survival of the insulating phase, because the energy $\ep$ of a localized single-particle state does not grow unbounded: hot particles quickly escape from the system.  In the thermodynamic limit, however, the exponential increase of the localization length leads to the disappearance of the insulating phase above a critical temperature $T_c$. %, related to the mean free path induced by the disorder. 
A similar conclusion was also obtained in Ref. \cite{Nandkishore} through another approach. An opposite situation for the thermodynamic limit is found in Ref. \cite{Gornyi}, where many-body localization is claimed to be unstable in any continuum system, irrespective of dimensionality, with a notable exception of one-dimensional Gaussian (white noise) disorder.% \cite{Aleiner}. 

In this paper, we discuss the stability of many-body localization in continuum two-dimensional (2D) systems. The 2D case shows a stronger (exponential) growth of the localization length with energy compared to one dimension. The nature of quantum statistics is not crucial for this problem, and we consider disordered bosons for convenience. First, we review the arguments of Refs. \cite{Nandkishore,Gornyi,Bertoli}, in which different criteria characterizing the MBLDT in continuum systems are employed. Then, building and improving on the results of Ref. \cite{Bertoli}, we rule out the arguments of \cite{Gornyi} on the absence of MBL in continuum systems of ultracold particles.
%we show that the arguments of \cite{Gornyi} may not be decisive enough to definitely rule out the possibility of MBL in continuum systems.

In general, the stability of the MBL phase is controlled by a parameter accounting for an increase of the phase space available for the transition when raising the temperature $T$. The key point is to compare the matrix element of interaction to the accessible level spacing \cite{Basko}. When this ratio exceeds a model-dependent value of order unity, then delocalization takes place. The parameter controlling many-body delocalization was derived  in Ref. \cite{Bertoli} on the basis of methods developed in Refs. \cite{Basko, Aleiner}. It is given by the probability $P_\al$ that for a given one-particle localized state $\ket\al$  there exist three other states $\ket{\al'},\ket{\be},\ket{\be'}$ for which the matrix element of interaction for the transition from the two-particle state $\ket{\al,\be}$ to $\ket{\al',\be'}$ exceeds the energy mismatch $\De_{\al\be}^{\al'\be'} = |\epsilon_{\alpha}+\epsilon_{\beta}-\epsilon_{\alpha'}-\epsilon_{\beta'}|$. For a short-range interaction $H_{\text{int}}$, one gets the probability:
\begin{equation}\label{probMBLDT}
P_\alpha= {\sum_{\al'\beta\beta'}}' \frac{\langle\alpha',\beta'|H_{\text{int}}|\alpha,\beta\rangle}{\Delta_{\alpha\beta}^{\alpha'\beta'}}\sim C,
\end{equation}
where $C$ is a parameter of order unity. %If $P_\al>C$ even for infinitesimal coupling, delocalization takes place.

In Ref. \cite{Bertoli}, the localized phase is protected by the high-energy truncation of the energy distribution function. On the contrary, if the particle energy grows unbounded, then delocalization takes place above a critical temperature $T_c$ that is interaction-independent, i.e. the insulating phase disappears even without interaction between particles. This conclusion, which apparently contradicts the commonly accepted Anderson localization of all single-particle eigenstates in 2D, was made and interpreted in Ref. \cite{Nandkishore}. The interpretation is based on the exponential increase of the localization length with energy. In order to estimate the ``conductivity'' one has to integrate the Bose distribution function multiplied by $\exp(-L/\zeta(\epsilon))$, where $L$ is the linear dimension of the system. Evaluating the integral by using the saddle point approximation, one obtains a power-like rather than exponential decrease of the ``conductivity'' with increasing $L$. The decrease that is slower than exponential can be interpreted as a disappearance of the insulating phase. The peculiarity of this rather academic problem follows from the fact that single-particle energies, which dominate the energy integral, increase logarithmically with $L$ and thus become infinite in the thermodynamic limit. For realistic systems, the exponential growth of the localization length is limited by e.g. a finite size of the system or, as in Ref. \cite{Bertoli}, by a truncation of the energy distribution function, and the 2D localization is restored. 

As noted, in Ref. \cite{Nandkishore} the author obtained the same result of an interaction-independent critical temperature, albeit with a different method. Namely, the MBLDT criterion in Ref. \cite{Nandkishore} contained the occupation number of the initial single-particle state. This statement is erroneous as noted in Refs. \cite{Gornyi,Bertoli}. However, the authors of Ref. \cite{Gornyi} concluded that MBL is unstable in continuum systems at any non-zero temperature, after rephrasing the MBLDT criterion as an energy exchange between ``hot'' and ``cold'' particles, i.e. particles with high and low energy, with intermediate energies playing no role.

%Within the framework of Ref. \cite{Bertoli}, t
As it is clear from Eq. \eqref{probMBLDT} (see, e.g. \cite{Bertoli}), the initial state single-particle occupation number does not enter the criterion for the MBLDT. Nevertheless, it was found \cite{Bertoli} that for a fixed interaction there exists a range of temperatures $T<T_c$ and disorder strengths, where the competition between the exponentials from the localization length (given by Eq. \eqref{eq:locLengthEn} below) and the distribution function $f(\ep)=(e^{(\ep-\mu)/T)}-1)^{-1}$ is ``won'' by the latter when increasing energy. An assumption adopted in Ref. \cite{Bertoli} is that the corresponding initial and final single-particle states are nearest neighbors in energy space, i.e. $\ep_\al\approx\ep_{\al'}$ and $\ep_\be\approx\ep_{\be'}$. The energies $\ep_\al$ and $\ep_\be$ may differ at will. Below, we relax this approximation in the criterion of delocalization. Before describing our results, let us briefly summarize the MBLDT criterion of Ref. \cite{Gornyi}. The key point of Ref. \cite{Gornyi} is that one should think of the whole system as containing two subsystems: ``hot'' and ``cold''  particles. Initially the cold particles act as a bath for the hot ones, creating delocalized excitations in the hot system, which in turn act as a bath for the cold system. This extends many-body delocalization over the whole spectrum, including intermediate states. From equations (1) and (9) of Ref. \cite{Gornyi}, we can write the delocalization parameter $\eta_{hc}$, which plays the same role as $C$ in our equation \eqref{probMBLDT}, as:
\beq \label{hotCold}
\eta_{hc} =\fr{V_{hc}N_{\text{eff}}}{\De_h}.
\eeq
Here $V_{hc}$ is the matrix element coupling two hot states with two cold states, $N_{\text{eff}}$ is the characteristic number of pairs in the cold system with which a hot particle can hybridize, and $\De_h$ is just the level spacing for the hot particles. Note that the structure of Eq. \eqref{hotCold} is similar to Eq. \eqref{probMBLDT}, but involves only coupling between hot and cold particles. The authors argue in favour of an effective $T$-dependent single-particle mobility edge so that states above such energy are always delocalized. This provides transport at any finite temperature, and no insulator is found other than the zero-temperature Bose glass \cite{Falco}.\\

Let us now address the question of the stability of the finite-temperature insulator in the context of the model introduced in Ref. \cite{Bertoli} for 2D interacting disordered bosons. The Hamiltonian of the system reads:
\beq
\hat H = \hat H_0+\hat H_{int},
\eeq
where
\beq
\hat H_0=\int \!d^2\!r\left(-\hat\Psi^\dagger(\textbf{r})\frac{\hbar^2}{2m}\nabla^2\hat\Psi(\textbf{r})+\hat\Psi^\dagger(\textbf{r})U(\textbf{r})\hat\Psi(\textbf{r})\right)\label{eq:HAnd},
\eeq
\beq
\hat H_{int} = g \!\int \!d^2\!r \;\;\hat\Psi^\dagger(\textbf{r})\hat\Psi^\dagger(\textbf{r})\hat\Psi(\textbf{r})\hat\Psi(\textbf{r}).\label{eq:HInt}
\eeq
Here $\hat\Psi(\textbf{r})$ are bosonic field operators. The first term in Eq. \eqref {eq:HAnd} is the kinetic energy of particles with mass $m$, while the second term accounts for the random potential $U(\tbf{r})$. The interparticle interaction, Eq. \eqref{eq:HInt}, describes a contact interaction between particles, with coupling constant $g>0$. We will consider the case of a disordered Gaussian short-range potential $U(\tbf{r})$ with zero mean, amplitude $U_0$, and correlation length $\si$. The energy and length scales of the disorder are \cite{Lifshitz,Zittartz}:
\begin{equation}
\epsilon_\ast=\frac{ mU_0^2\sigma^2}{\pi\hbar^2}; \qquad \zeta_\ast = \sqrt{\frac{2e^2}{\pi}}\frac{\hbar^2}{mU_0\sigma}.
\end{equation}
In two dimensions, single-particle states are localized with a localization length depending exponentially on the energy%, which was observed also in \cite{Manai}
. The single-particle localization length in two dimensions at $\epsilon>\epsilon_\ast$ can be written in the form  \cite{Lee}:
\begin{equation} \label{eq:locLengthEn}
\zeta(\epsilon)=\frac{\zeta_\ast}{e} \sqrt{\frac{\epsilon}{\epsilon_\ast}}\; \exp\left(\frac{\epsilon}{ \epsilon_\ast}\right),
\end{equation}
so that $\zeta(\epsilon_\ast)=\zeta_\ast$. 
At energies $|\epsilon|\lesssim\epsilon_\ast$, the energy dependence of $\zeta$ is weak, and one can approximate the localization length as $\zeta(\epsilon)\approx\zeta_\ast$. The density of states (DoS) for clean 2D bosons in the continuum is energy independent, $\rho_0 = m/2\pi\hbar^2$. The presence of the disordered potential creates the so-called Lifshitz tails, negative-energy states with a DoS decaying exponentially with increasing the absolute value of the energy \cite{Lifshitz,Zittartz}. We approximate $\rho(\epsilon)\simeq\rho_0$ for energies $\ep\ge-\eps$, omitting exponentially small values of the DoS for $\ep<-\eps$.
Besides that, we consider the weakly interacting regime, with a small parameter given by:
\begin{equation}    \label{sp}
\frac{ng}{T_d}=\frac{mg}{2\pi\hbar^2}\ll 1.
\end{equation}
Here $T_d=2\pi\hbar^2n/m$ is the degeneracy temperature, with $n$ being the mean density. Finally, we assume weak disorder, so that 
\begin{equation}     \label{wd}
\epsilon_\ast\ll T_d.
\end{equation}

What should happen when a hot localized particle with energy $\epsilon$ and localization length $\zeta(\epsilon)$ is present in the system, and it interacts with a cold cloud? A rough estimate similar to Eq. \eqref{hotCold} suggests  that the matrix element $V_{hc}$ in this case is proportional to $\sim g/\zeta^2(\epsilon)$. The level spacing is just $\De_h\sim1/\rho_0 \zeta^2(\epsilon)$. For particles of the cloud which have energies approaching $\epsilon$, the effective number of channels is $N_{\text{eff}}\sim n \zeta^2(\epsilon)\exp(-\epsilon/T)$. Putting this altogether we obtain: 
\beq
\eta_{hc} \sim \frac{n^2g\zeta^2_\ast}{e^2T_d} \frac{\epsilon}{\epsilon_\ast}\exp\left[-\epsilon\left(\frac{1}{T}-\frac{2}{\epsilon_\ast}\right)\right].
\eeq
Accordingly, it is possible that $\eta_{hc}\lesssim 1$ provided that $T<\epsilon_{\ast}/2$, which means that there is an insulating phase at these temperatures in the thermodynamic limit, in agreement with Ref. \cite{Bertoli}. However, the finite-temperature insulator might be merely a consequence of our approximations. Indeed, in Ref. \cite{Gornyi} the question is addressed by inserting the effective mobility edge, which provides the system with an effective conduction band populated by many delocalized excitations.
%Putting this altogether with $\eta_{hc}\sim 1$, we get:
%\beq
%\fr{mg}{\hbar^2}\sim\fr{1}{n\ze^2(T)}.
%\eeq
%This equation gives the localization-delocalization transition at a given temperature and disorder strength. However, the finite-temperature insulator might be merely a consequence of our crude approximation, where we considered a single hot particle and not a collection of them. Indeed, in Ref. \cite{Gornyi} the question is addressed by inserting the effective mobility edge, which provides the system with an effective conduction band populated by many delocalized excitations.

We now go back to the MBLDT criterion given by Eq. \eqref{probMBLDT}, which takes the form:
\begin{equation} \label{probMBLDT2}
P_\alpha= {\sum_{\al'\beta\beta'}}' \frac{\langle\alpha',\beta'|H_{int}|\alpha,\beta\rangle}{\Delta_{\alpha\beta}^{\alpha'\beta'}}={\sum_{\al'\beta\beta'}}' \frac{U_{\alpha\beta}^{\alpha'\beta'}N_{\alpha\beta}^{\alpha'\beta'}}{\Delta_{\alpha\beta}^{\alpha'\beta'}}\sim C,
\end{equation}
where:
\alig{
&N_{\alpha\beta}^{\alpha'\beta'}=\overline{\sqrt{|N_\beta(1+N_{\alpha'})(1+N_{\beta'})-N_{\alpha'}N_{\beta'}(1+N_\beta)|}}\label{channels}\\ 
&U_{\alpha\beta}^{\alpha'\beta'}=g\int \psi_\alpha(\tbf{r})\psi_{\alpha'}(\tbf{r})\psi_\be(\tbf{r})\psi_{\be'}(\tbf{r})d^2\!r \label{matrElU}\\
&\De_{\al\be}^{\al'\be'} = |\epsilon_{\alpha}+\epsilon_{\beta}-\epsilon_{\alpha'}-\epsilon_{\beta'}|. \label {levspecfull}
}
The prime in the summation in Eq. (\ref{probMBLDT2}) means that we are summing over a length scale of the localization length (the lowest one among the four states). The factor $N_{\alpha\beta}^{\alpha'\beta'}$ accounts for the number of possible (direct and inverse) processes $\al,\be\leftrightarrow\al',\be'$ involving a given state $\al$, and $N_{\al'},N_\be,N_{\be'}$ are the occupation numbers (not the averages, and the average is only taken for the square root expression in Eq. \eqref{channels}). Following Ref. \cite{Basko}, we %do not consider interference terms and
 replace the difference of square roots with the square root of the difference. The quantity $\psi_\alpha(\tbf{r})$ is the one-body wavefunction of a localized state and the bar in the right hand side  of \eqref{channels} means average value. While the true form of the wavefunction is given by an exponentially decaying wavepacket, we make the following approximation:
\beq
\psi_\alpha(\tbf{r})=
\begin{cases}
\ze_\al^{-1}&\quad |\tbf{r}-\tbf{r}_\alpha| < \ze_\al/2.\\
0&\text{otherwise.}
\end{cases}
\eeq
%This means that the integral in Eq. \eqref{matrElU} is performed over a disk of radius $\zeta_{\text{min}}$. 
Our main goal is to demonstrate the stability of the MBL state. In order to do this we neglect the well known (see, e.g. \cite{Mirlin}) dependence of the matrix element of the two-body $\alpha,\beta\rightarrow\alpha',\beta'$ scattering, 
$U_{\alpha\beta}^{\alpha'\beta'} $, on the energy differencies and estimate it as
\beq
U_{\alpha\beta}^{\alpha'\beta'} \approx 
 g \fr{\min(\ze_\al^2,\ze_{\al'}^2,\ze_\be^2,\ze_{\be'}^2)}{\ze_\al \ze_{\al'}\ze_\be \ze_{\be'}} 
\eeq 
%We have neglected here the fall-off of the matrix element with increasing the energy difference \cite{Mirlin}. %because the localization length dominates, being an exponential function of energy. 
%It is instructive to compare Eq. \eqref{probMBLDT2} with Eq. \eqref{hotCold}, used in Ref. \cite{Gornyi}. One may identify the state $\al$ with the initial hot state decaying into the three-particle state given by another hot ($\al'$) and two cold ($\be,\be'$) particles. %In this context, the approximation taken in Ref. \cite{Bertoli} of pairwise energy neighbors should still hold.\\ 

Let us now look at equation \eqref{channels}. The quantities $N_{\al'},N_{\be},N_{\be'}$ are actually integers representing the occupation of the state in Fock space. If the corresponding average values $\overline N_{\al'},\overline N_{\be},\overline N_{\be'}$ are large, then fluctuations are small and we may substitute the average values of the occupation numbers in the r.h.s. of Eq.  \eqref{channels}. Assuming that energies $\ep$ are almost equal to each other pairwise, i.e. $\epsilon_{\alpha}\approx\epsilon_{\alpha'}$ and $\epsilon_{\beta}\approx\epsilon_{\beta'}$, one then recovers the expression $N_{\alpha\beta}^{\alpha'\beta'}\approx\overline N_\be$ as in Ref. \cite{Bertoli}. Going beyond this approximation involves the calculation for each distinct case when some of the average occupation numbers are small. Detailed calculations are given in the Appendix. It turns out that the MBL state is satable if $g<g_c(\epsilon_{\alpha})$, where  
\aligns{
&g_c(\ep_\al)=C\left(\sum_{\be,\al',\be'>\al} \frac{N_{\alpha\beta}^{\alpha'\beta'}}{\De_{\al\be}^{\al'\be'} } \frac{\ze_\al}{\ze_{\al'}\ze_\be\ze_{\be'}}+\!\!\!\!\!\!\!\!\sum_{\al'<\al,\be,\be'} \frac{N_{\alpha\beta}^{\alpha'\beta'}}{\De_{\al\be}^{\al'\be'} } \frac{\ze_{\al'}}{\ze_{\al}\ze_\be\ze_{\be'}}\right.\\
&\!\!\!+\!\!\!\!\!\sum_{\be<\al,\al',\be'} \frac{N_{\alpha\beta}^{\alpha'\beta'}}{\De_{\al\be}^{\al'\be'} } \frac{\ze_\be}{\ze_{\al'}\ze_\al\ze_{\be'}}+\!\!\!\!\!\!\!\!\left.\sum_{\be'<\al,\al',\be} \frac{N_{\alpha\beta}^{\alpha'\beta'}}{\De_{\al\be}^{\al'\be'} } \frac{\ze_{\be'}}{\ze_{\al'}\ze_\be\ze_{\al}}\right)^{-1}\!\!\!\!\!\!\!\!.\numberthis\label{eq:newCrit}
}
For the MBL state to remain stable the condition $g<g_c(\epsilon_{\alpha})$ should be satisfied for all $\alpha$, i.e. the condition of the stability is
\begin{equation}     \label{stabcr}
g<g_c={\rm min}\{g(\epsilon_{\alpha})\}.
\end{equation}

According to Eq. (\ref{eq:newCrit}), the critical coupling $g_c(\epsilon_{\alpha})$ is determined by the typical smallest value of the energy mismatch. In order to estimate ${\rm min}\{\Delta_{\alpha\beta}^{\alpha'\beta'}\}$ relying on Eq. (\ref{levspecfull}) note that all four states $\alpha,\beta,\alpha',\beta'$ should be localized nearby. The nearest neighbour spacing between such states can be estimated as $\delta=(\rho_0\zeta^2)^{-1}$, where $\rho_0$ is the density of states. Since $\zeta$ depends on the energy, one has $\delta=\delta(\epsilon)$. Therefore, we have    
\beq \label{levspacMin}
{\rm min}\{\De_{\al\be\;\min}^{\al'\be'}\} = \min \{\de_{\al'},\de_\be,\de_{\be'}\},
\eeq
where $\delta_{\beta}\equiv\delta(\epsilon_{\beta})$ and the same for $\alpha'$, $\beta'$. It should be noted that the assumption of the energy-independence of the matrix element $U_{\alpha\beta}^{\alpha'\beta'}$, which we adopted, substantially reduces the estimate for  ${\rm min}\{\De_{\al\be\;\min}^{\al'\be'}\}$ compared with Ref. \cite{Bertoli}: the requirement of \cite{Bertoli} that $\alpha$ and $\alpha'$, as well as $\beta$ and $\beta'$ should be nearest neighbours in energy leads to ${\rm min}\{\De_{\al\be\;\min}^{\al'\be'}\}={\rm max}(\delta_{\alpha},\delta_{\beta})$. Including the possibility of a smaller denominator in Eq. \eqref{eq:newCrit} can favour delocalization.

Below we solve Eq. \eqref{eq:newCrit} numerically and first check whether the MBL phase exists in the low-temperature limit. As $N_{\alpha\beta}^{\alpha'\beta'}$ depends on the chemical potential $\mu$, we establish a relation between $\mu$, $n$, and $T$ from the normalization condition
\beq \label{numbeq}
n=\int_{-\eps}^\infty\rho_0 \overline N_\ep d\ep.
\eeq\\

In order to calculate the critical coupling $g_c$ and its temperature dependence, one has to evaluate the occupation numbers of single-particle states. Following Ref. \cite{Michal}, we write an expression for the energy corresponding to the configuration $\{N_\al\}$ of the occupation numbers as:
\beq %n(n-1)/2 is number of pairs?
E(\{N_\al\}) = \sum_\al (\ep_\al N_\al + g N_\al (N_\al -1)/2\ze_\al^2).
\eeq
The grand canonical partition function becomes:
\beq
Z=\prod_\al Z_\al,
\eeq
where:
\beq \label{partF}
Z_\al = \sum_{n=0}^\infty \exp (-((\ep_\al - \mu)n +gn(n-1)/2\ze_\al^2)/T).
\eeq
For a large average occupation number $\overline N_\al\gg 1$, fluctuations are small. One linearizes the exponent around $\overline N_\al$ and the partition function reads:
\beq
Z_\al \approx \frac{1}{1-\exp (-(\ep_\al - \mu +g\overline  N_\al/\ze_\al^2)/T ) }.
\eeq
Dropping the index $\alpha$, we have the following expression for the average occupation numbers of single-particle states on the insulator side:
\begin{equation}     \label{Negen}
\overline N_{\epsilon}=T\pder{\ln Z}{\mu}\approx\left[\exp\left(\frac{\epsilon-\mu+\overline N_{\epsilon}g/\zeta^2(\epsilon)}{T}\right)-1\right]^{-1}.
\end{equation}
For $\overline N_{\epsilon}\gg 1$ we expand the exponent in Eq. (\ref{Negen}) and obtain: 
\begin {equation}     \label{Nelarge}
\overline N_{\epsilon}=\frac{\zeta^2(\epsilon)}{2g}\left(\mu-\epsilon +\sqrt{(\mu-\epsilon)^2+\frac{4Tg}{\zeta^2(\epsilon)}}\right).
\end{equation}
For small average occupation numbers $\overline N_\ep \ll 1$ we neglect the interaction term in the exponent of \eqref{partF}.  This gives the Boltzmann distribution $\overline N_\ep \approx e^{-(\epsilon-\mu)/T} $.

Below, we set the value of the constant $C=1$. At $T=0$, Eq. \eqref{Nelarge} gives
\begin {equation}     \label{NeTzero}
\overline N_{\epsilon}=\frac{\zeta^2(\epsilon)(\mu-\epsilon)}{g}\theta(\mu-\epsilon),
\end{equation}
where $\theta(\mu-\epsilon)$ is the Heaviside theta function. Combining equations (\ref{eq:newCrit}), (\ref{levspacMin}), and (\ref{NeTzero}) we find a critical disorder:
\beq
\eps(0)^{MBL}=0.87 ng.
\eeq
The corresponding chemical potential is $\mu=1.64ng$. This result is in good agreement with the one from Ref. \cite{Bertoli}, as well as with the microscopic analysis of tunneling between bosonic lakes \cite{Falco}. The critical disorder is higher than the one in Ref. \cite{Bertoli}. This is expected, as we include smaller level spacing than before and more highly energetic processes.

At temperatures $T\ll\epsilon_\ast^2/T_d=m\epsilon_{\ast}^2/2\pi\hbar^2n$, the  critical disorder is practically temperature independent. In the temperature interval, $\epsilon_\ast^2/T_d\ll T\ll \epsilon_\ast$, in the thermodynamic limit the average occupation number $\overline N_{\epsilon}$ is large when $\ep<\mu$. This is because the chemical potential decreases with increasing $T$, and becomes negative when $T$ is a fraction of $\eps$ \cite{Bertoli}. One integrates Eq. \eqref{eq:newCrit} with Eq. \eqref{numbeq} using average occupation numbers given by (see also Refs. \cite{Michal,Bertoli}):

 \begin{equation}\label{eq:occNumber}
\overline N_\epsilon = \begin{cases}
\dfrac{\zeta^2(\ep)}{2g} \left(\mu-\epsilon+\sqrt{\strut(\epsilon-\mu)^2+\fr{4Tg}{\zeta^2(\ep)}}\;\right)&;\quad\epsilon<\mu\\[1em]
e^{-(\epsilon-\mu)/T} &;\quad\mu<\epsilon.
\end{cases}
\end{equation}
 
Note that at this stage we do not consider a truncation in the energy distribution.
 
\begin{figure}[h!]
    \centering
        \includegraphics[width=0.45\textwidth]{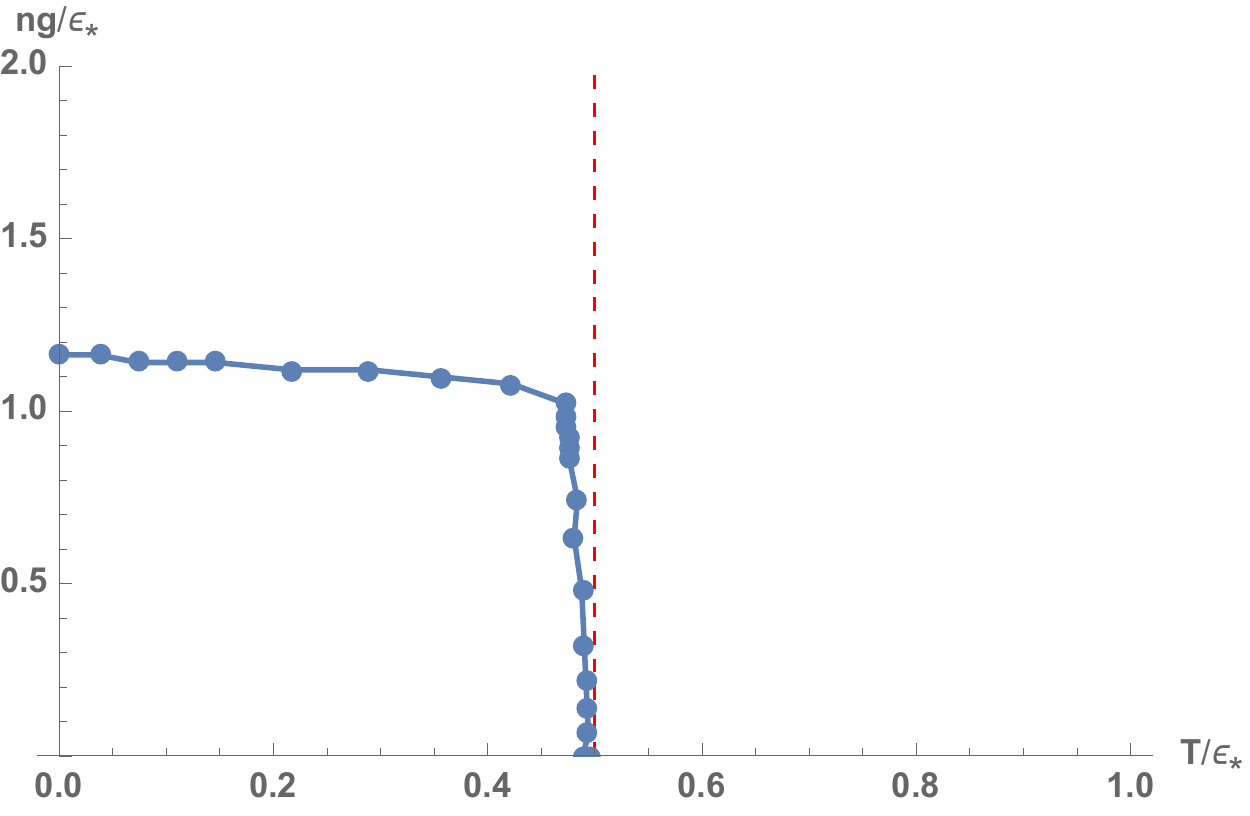}
\caption
{The critical coupling $ng_c/\eps$ as a function of temperature $T/\eps$ obtained by numerically solving equations \eqref{eq:newCrit} and \eqref{numbeq}, without a truncation in the energy distribution function. The blue dots represent the values of $T$ for which we solved the equations. The critical coupling tends to zero at $T\approx \eps/2$. The dashed red line indicates $T=\eps/2$. Each of the dots is obtained with a numerical uncertainty of not more than 5\%. 
}
\label{fig:ngHC}
\end{figure}

\begin{figure}[h!]
    \centering
        \includegraphics[width=0.45\textwidth]{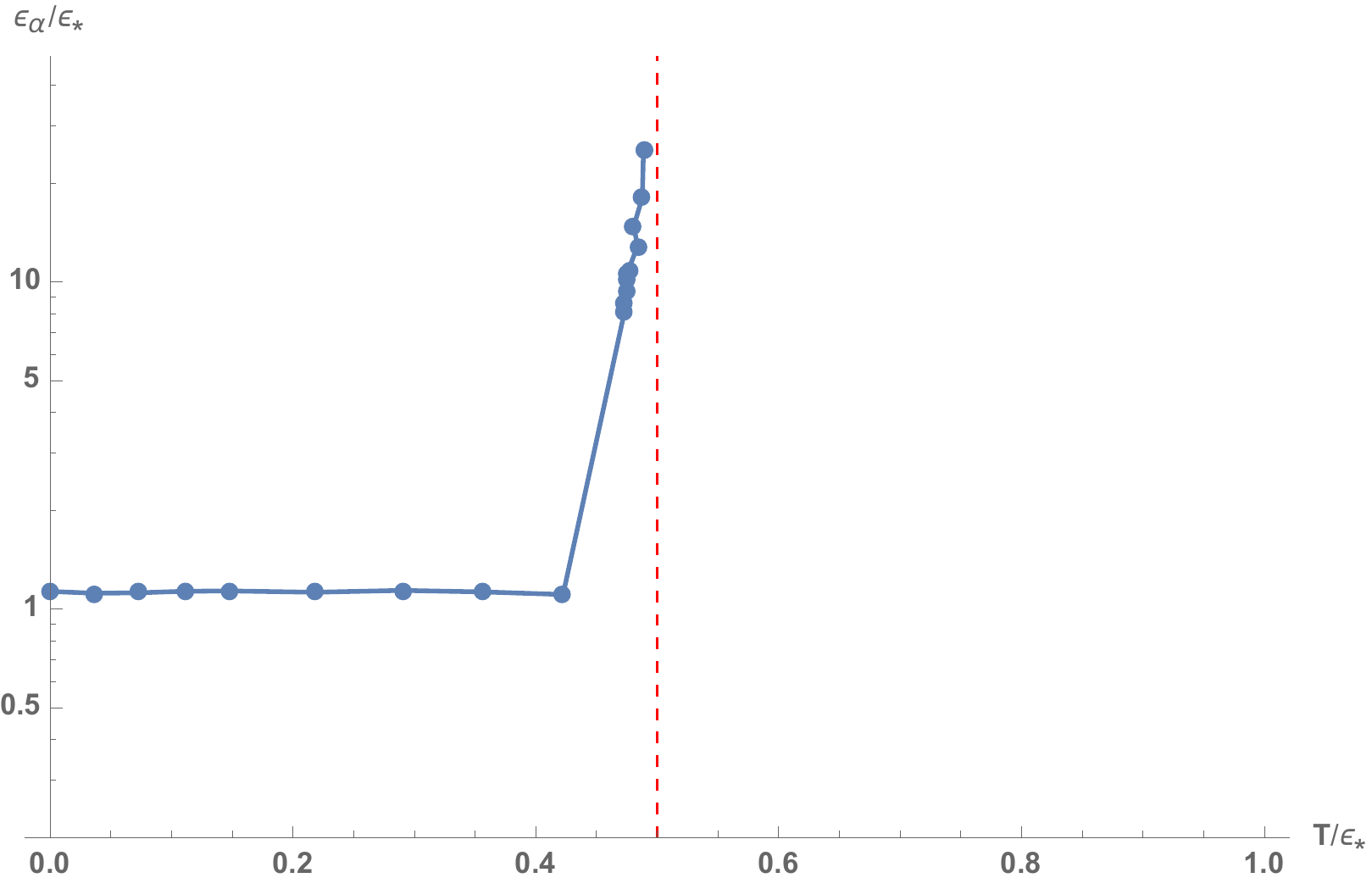}
\caption
{The value of the energy $\ep_\al/\eps$ plotted versus the temperature $T/\eps$, in a semi-logarithmic plot. The numerical solutions, given by the blue points, are linearly interpolated. The dashed red line is $T=\eps/2$. Between $T\approx0.42\eps$ and $T\approx0.47\eps$, the value of $\ep_\al$ jumps to high energies.}
\label{fig:alpha}
\end{figure}

Fig. \ref{fig:ngHC} shows the obtained results. Remarkably, the physical picture resulting from Ref. \cite{Bertoli} survives in the presence of a number of processes that were not taken into account there. For very low temperatures $T\ll \eps/2$, the insulator is stable and the critical coupling is only slightly reduced by an increase in temperature. Most importantly, delocalization is driven in this regime by the low-energy states, as we see from Fig. \ref{fig:alpha}, where we plot the value of $\ep_\al$ as a function of $T$. Indeed, we may identify the value of $\ep_\al$ as the energy of the typical states that cause delocalization through the interaction. Finding a low value of $\ep_\al$ for low temperatures implies that the resonant subnetwork in Fock space typically involves states at low energies. \\
Hot particles start to dominate only when $T\approx0.47\eps$. Then, delocalization takes place as a result of hybridization of high-energy particles, decaying into three-body excitations. This is compatible with the picture of a hot-cold mixture proposed in Ref. \cite{Gornyi} and was already noted in Ref. \cite{Bertoli} when looking at the behavior of $\ep_\al$ in the thermodynamic limit.

Let us remark that in Eq. \eqref{eq:newCrit} we take into account all processes that are resonant irrespective of their energies. The only assumption is that two highly energetic states may be taken as energy neighbors when they interact with two cold states, which is consistent with the analysis in the hot-cold mixture. Nevertheless, the obtained results show that localization is present in the low-temperature regime. This is in direct contrast with the proposal of considering the hot and cold subsystems as two separated entities that act as a bath for one another. %While fascinating, might be missing the important point that not all transitions are equiprobable, something taken into account when formulating the MBLDT criterion as in Eq. \eqref{probMBLDT}.
However, it may well be that this is the situation when $T\to\eps/2$ and beyond.

\begin{figure}[h!]
    \centering
        \includegraphics[width=0.45\textwidth]{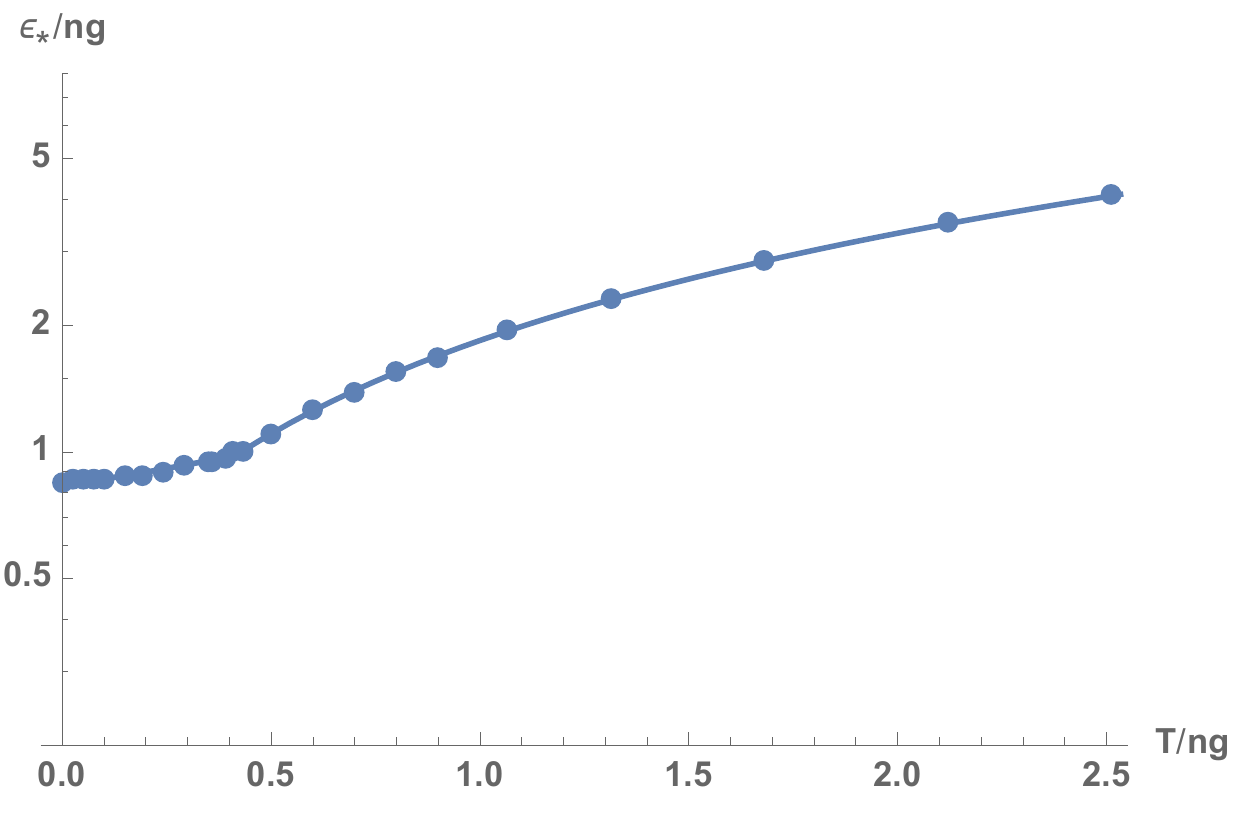}
\caption
{The MBLDT in the presence of a truncated energy distribution. The energy barrier is at $\ep_b = 1.64ng+5T$. We used $T_d/ng=10$.
}
\label{fig:trunc}
\end{figure}

To complete our analysis, we introduce a truncation in the energy distribution function. The value of the truncation energy $\ep_b$ for cooling to temperatures $T\lesssim ng$ is typically equal to $ng+\eta T$, with $\eta$ varying between 5 and 8 \cite{walraven, ketterle}. We match therefore the zero-temperature result by setting $\ep_b = 1.64ng+\eta T$, and below we use $\eta=5$. Increasing the value of $\eta$ does not change significantly the physical picture. The result is in good agreement with the one in Ref. \cite{Bertoli}, as shown in Fig. \ref{fig:trunc}. Even though it is slightly reduced with respect to the phase diagram presented in Ref. \cite{Bertoli}, the insulating phase survives at all temperatures.

It is actually not surprising that the behavior of the system is almost the same, even after relaxing some of the approximations made in Ref. \cite{Bertoli}. As we already remarked, very hot particles are not present and cannot drive delocalization. 

%We only looked at the case of a weak Gaussian white-noise disorder. A different disorder correlation function can in principle alter the shape of the phase diagram, because a finite disorder correlation length implies a stronger growth %of the localization length with energy. However, this should not change the important result that, in the presence of a truncated distribution function, the insulator phase is present at any temperature. In this respect, our result is not %dependent on the microscopic details of the disorder.

In conclusion, we see that for short-range interacting bosons the MBL phase is stable at temperatures below a critical disorder-dependent value in spite of the presence of processes involving highly energetic particles. %It should be noted that this does not automatically mean that MBL is present, as we did not consider other phenomena such as spectral diffusion and more complex interactions \cite{gornyi2017spectral}. 
%We thus have provided further arguments for the stability of the MBL phase in the continuum, strengthening thus the conclusions made in Ref. \cite{Bertoli}.  

The research leading to these results has received funding from the European Research Council under European Community's Seventh Framework Programme (FP7/2007-2013 Grant Agreement no.341197).

\newpage

\onecolumngrid
\section*{Appendix}
We need to calculate the average value of the square root
\beq \label{eq:channelsavg}
N_{\alpha\beta}^{\alpha'\beta'}=\overline{\sqrt{|N_\beta(1+N_{\alpha'})(1+N_{\beta'})-N_{\alpha'}N_{\beta'}(1+N_\beta)|}}.
\eeq
When the average occupation numbers are large, we substitute them directly into equation \eqref{eq:channelsavg} because fluctuations are small. We get:
\beq \label{channels1}
\!N_{\alpha\beta}^{\alpha'\beta'}\!= \!\sqrt{|\overline N_\beta(1+\overline N_{\alpha'})(1+\overline N_{\beta'})-\overline N_{\alpha'}\overline N_{\beta'}(1+\overline N_\beta)|} \!=\!\sqrt{|\overline N_\beta\overline N_{\alpha'}+\overline N_\beta\overline N_{\beta'}+\overline N_\beta-\overline N_{\alpha'}\overline N_{\beta'}|} \!\quad;\!\quad  \overline N_\be, \overline N_{\be'},  \overline N_{\al'}  \gg 1,
\eeq
where the average occupation number is given by Eq, \eqref{eq:occNumber}. Similarly, we keep the largest terms when one of the average occupation numbers is small:
\alig{
N_{\alpha\beta}^{\alpha'\beta'}= \sqrt{\overline N_{\alpha'}\overline N_{\beta'}} \qquad;\qquad  \overline N_{\be'},  \overline N_{\al'} \gg 1; \overline N_\be \ll 1 \\
N_{\alpha\beta}^{\alpha'\beta'}= \sqrt{\overline N_{\alpha'}\overline N_{\beta}} \qquad;\qquad  \overline N_{\be},  \overline N_{\al'} \gg 1; \overline N_{\be'} \ll 1 \\
N_{\alpha\beta}^{\alpha'\beta'}= \sqrt{\overline N_{\beta}\overline N_{\beta'}} \qquad;\qquad  \overline N_{\be'},  \overline N_{\be} \gg 1; \overline N_{\al'}\ll 1.
}
When two of the average occupation numbers are small, we omit the interparticle interaction for these states and calculate the probability of having $j$ particles in the state with energy $\ep$ as:
\beq \label{probBoltzAvg}
p_j = (1-e^{-(\ep-\mu)/T})e^{-(\ep-\mu)j/T}.
\eeq
Let us look first at the case where $\overline N_{\al'} \gg 1$ and $ \overline N_{\be'},  \overline  N_\be \ll 1$. We then have:
\beq
N_{\alpha\beta}^{\alpha'\beta'}= \sum_{N_\be = 0}^\infty \sum_{N_{\be'} = 0}^\infty p_{N_\be} p_{N_{\be'}}\sqrt{|\overline N_{\al'}(N_{\be}-N_{\be'})|}.
\eeq
The main contribution comes from the terms with
\beq
N_\be = 1, N_{\be'} = 0\qquad;\qquad N_\be = 0, N_{\be'} = 1.
\eeq
Taking into account that $e^{-(\ep_{\be(\be')}-\mu)/T}\ll1$, this yields:
%The cases with higher occupation numbers $N_\be\ge 2$ and the case $N_\be = N_{\be'}=1$ can be neglected because the exponential term in \eqref{probBoltzAvg} is very small. Then, from Eq. \eqref{eq:channelsavg} we get:
%\aligns{
%N_{\alpha\beta}^{\alpha'\beta'} \approx & \sqrt{1+\overline N_{\al'}}(1-e^{-(\ep_\be-\mu)/T})e^{-(\ep_\be-\mu)/T}(1-e^{-(\ep_{\be'}-\mu)/T})\\
%&+\sqrt{\overline N_{\al'}}(1-e^{-(\ep_{\be'}-\mu)/T})e^{-(\ep_{\be'}-\mu)/T}(1-e^{-(\ep_{\be}-\mu)/T}). \numberthis}
%Using the expression for small average occupation numbers
%\beq
%\overline N_\ep \approx e^{-(\ep-\mu)/T}(1-e^{-(\ep-\mu)/T})
%\eeq
%we find:
\aligns{
N_{\alpha\beta}^{\alpha'\beta'} = &\sqrt{\overline N_{\al'}}(1-e^{-(\ep_\be-\mu)/T})e^{-(\ep_\be-\mu)/T}(1-e^{-(\ep_{\be'}-\mu)/T}) \\
&\qquad+ \sqrt{\overline N_{\al'}}(1-e^{-(\ep_{\be'}-\mu)/T})e^{-(\ep_{\be'}-\mu)/T}(1-e^{-(\ep_{\be}-\mu)/T}) \\
\approx& \sqrt{\overline N_{\alpha'}} (\overline N_{\beta}+\overline N_{\beta'}),\numberthis
}
where $\overline N_{\beta'}= e^{-(\ep_{\be'}-\mu)/T}$ and $\overline N_{\beta}=e^{-(\ep_{\be}-\mu)/T}$. For small average occupation numbers $\overline N_{\beta}$ and $\overline N_{\beta'}$, the states $\be$ and $\be'$ have large energies. We take now the approximation $\ep_\be\approx\ep_{\be'}$ to find
\beq \label{approxNLE}
N_{\alpha\beta}^{\alpha'\beta'}\approx 2\overline N_{\beta}\sqrt{\overline N_{\alpha'}} \qquad;\qquad   \overline N_{\al'} \gg 1; \overline N_{\be'},  \overline  N_\be \ll 1.
\eeq
Similar calculations give:
\alig{
N_{\alpha\beta}^{\alpha'\beta'}\approx 2\overline N_{\alpha'}\sqrt{2\overline N_{\beta}} \qquad;\qquad  \overline N_{\be}\gg 1; \overline N_{\al'} ,  \overline  N_{\be'} \ll 1 \\
N_{\alpha\beta}^{\alpha'\beta'}\approx 2\overline N_{\beta}\sqrt{\overline N_{\be'}}\qquad;\qquad  \overline N_{\be'} \gg 1; \overline N_{\be}  \overline N_{\al'}\ll 1.
}\\
When all three of the average occupation numbers are small, using the same method we have:
\beq
N_{\alpha\beta}^{\alpha'\beta'} = \overline N_\be.
\eeq
The value of $\ep_\al$ must be chosen in such a way that off-resonant processes are avoided. As in Eq. \eqref{approxNLE}, we set the two highest energy equal to each other (when the average occupation numbers are small) in order to account for the fall-off of the matrix element when one of the energies becomes very high. Using equations \eqref{matrElU}-\eqref{levspecfull}, this gives the following MBLDT criterion in the thermodynamic limit:
\aligns{
&g \int_{\ep_\al}^\mu d\ep_{\al'} \int_{\ep_\al}^\mu d\ep_{\be'} \int_{\ep_\al}^\mu d\ep_{\be} \rho_0^3\fr{\sqrt{|\overline N_\beta(1+\overline N_{\alpha'})(1+\overline N_{\beta'})-\overline N_{\alpha'}\overline N_{\beta'}(1+\overline N_\beta)|}}{\De_{\al\be}^{\al'\be'}}\ze_\al\ze_\be\ze_{\al'}\ze_{\be'}+\\&
g \int_{-\eps}^{\ep_\al} d\ep_{\al'} \int_{\ep_{\al'}}^\mu d\ep_{\be'} \int_{\ep_{\al'}}^\mu d\ep_{\be} \rho_0^3\fr{\sqrt{|\overline N_\beta(1+\overline N_{\alpha'})(1+\overline N_{\beta'})-\overline N_{\alpha'}\overline N_{\beta'}(1+\overline N_\beta)|}}{\De_{\al\be}^{\al'\be'}}\fr{\ze_\be\ze^3_{\al'}\ze_{\be'}}{\ze_\al}+\\&
g \int_{-\eps}^{\ep_\al} d\ep_{\be} \int_{\ep_{\be}}^\mu d\ep_{\be'} \int_{\ep_{\be}}^\mu d\ep_{\al'} \rho_0^3\fr{\sqrt{|\overline N_\beta(1+\overline N_{\alpha'})(1+\overline N_{\beta'})-\overline N_{\alpha'}\overline N_{\beta'}(1+\overline N_\beta)|}}{\De_{\al\be}^{\al'\be'}}\fr{\ze_\be^3\ze_{\al'}\ze_{\be'}}{\ze_\al}+\\&
g \int_{-\eps}^{\ep_\al} d\ep_{\be'} \int_{\ep_{\be'}}^\mu d\ep_{\be} \int_{\ep_{\be'}}^\mu d\ep_{\al'} \rho_0^3\fr{\sqrt{|\overline N_\beta(1+\overline N_{\alpha'})(1+\overline N_{\beta'})-\overline N_{\alpha'}\overline N_{\beta'}(1+\overline N_\beta)|}}{\De_{\al\be}^{\al'\be'}}\fr{\ze_\be\ze_{\al'}\ze_{\be'}^3}{\ze_\al}+\\&
g \int_{\mu}^\infty d\ep_{\be} \int_{\ep_{\al}}^\mu d\ep_{\al'} \rho_0^2\fr{2\overline N_{\beta}\sqrt{\overline N_{\alpha'}}}{\De_{\al\be}^{\al'\be'}}\ze_\al\ze_{\al'}+g \int_{\mu}^\infty d\ep_{\be} \int_{\ep_{\al}}^\mu d\ep_{\be'} \rho_0^2\fr{2\overline N_{\beta}\sqrt{\overline N_{\be'}}}{\De_{\al\be}^{\al'\be'}}\ze_\al\ze_{\be'}+\\&
g \int_{\ep_{\al'}}^\mu d\ep_{\be} \int_{-\eps}^\mu d\ep_{\al'} \rho_0^2\fr{\sqrt{\overline N_{\alpha'}\overline N_{\beta}}}{\De_{\al\be}^{\al'\be'}}\fr{\ze_\be\ze^3_{\al'}}{\ze_\al^2}+g \int_{-\eps}^{\ep_\al} d\ep_{\al'} \int_\mu^\infty d\ep_{\be} \rho_0^2\fr{2\overline N_{\beta}\sqrt{\overline N_{\al'}}}{\De_{\al\be}^{\al'\be'}}\fr{\ze_\be\ze^3_{\al'}}{\ze_\al^2}+\\&
g \int_{-\eps}^\mu d\ep_{\be} \int_{\ep_\be}^\mu d\ep_{\al'} \rho_0^2\fr{\sqrt{\overline N_{\alpha'}\overline N_{\beta}}}{\De_{\al\be}^{\al'\be'}}\fr{\ze^3_\be\ze_{\al'}}{\ze_\al^2}+g \int_{-\eps}^{\mu} d\ep_{\be} \int_{\ep_{\be}}^\mu d\ep_{\be'} \rho_0^2\fr{\sqrt{\overline N_{\be'}\overline N_{\beta}}}{\De_{\al\be}^{\al'\be'}}\fr{\ze^3_\be\ze_{\be}}{\ze_\al^2}+\\&
g \int_{-\eps}^\mu d\ep_{\be'} \int_{\ep_{\be'}}^\mu d\ep_{\be} \rho_0^2\fr{\sqrt{\overline N_{\be'}\overline N_{\beta}}}{\De_{\al\be}^{\al'\be'}}\fr{\ze^3_{\be'}\ze_{\be}}{\ze_\al^2}+g \int_{-\eps}^{\ep_\al} d\ep_{\be'} \int_{\mu}^\infty d\ep_{\be} \rho_0^2\fr{2\overline N_{\beta}\sqrt{\overline N_{\be'}}}{\De_{\al\be}^{\al'\be'}}\fr{\ze^3_{\be'}\ze_{\be}}{\ze_\al^2}+\\&
3g\rho_0 \int_{\ep_\al}^\infty d\ep_{\be} \fr{\overline N_{\be}}{\De_{\al\be}^{\al'\be'}}+g\rho_0 \int_{\mu}^{\ep_\al} d\ep_{\be} \fr{\overline N_{\be}}{\De_{\al\be}^{\al'\be'}}\fr{\ze^2_{\be}}{\ze_\al^2}=C.}
We have taken the average occupation number to be large at energies smaller than the chemical potential, and checked that this is a good approximation.


\begin{thebibliography}{99}

\bibitem{Anderson}
P.W. Anderson, Phys. Rev.{\bf 109}, 1492 (1958).

\bibitem{50years}
E. Abrahams (Ed.), \textit{50 years of Anderson Localization}, World Scientific (vol. 26, 2010).

\bibitem{Fleishman}
L. Fleishman and P.W. Anderson, Phys. Rev. B {\bf 21}, 2366 (1980).

\bibitem{Altshuler}
B.L. Altshuler, Yu. Gefen, A. Kamenev, and L.S. Levitov, Phys. Rev. Lett. {\bf 78}, 2803 (1997).

\bibitem{Basko}
D.M. Basko, I.L. Aleiner, and B.L. Altshuler, Annals of Physics {\bf 321}, 1126 (2006).

\bibitem{Gornyi1}
I.V. Gornyi, A.D. Mirlin and D.G. Polyakov, Phys. Rev. Lett. {\bf 95}, 206603 (2005).

\bibitem{ReviewHuse}
R. Nandkishore and D. A. Huse, Annu. Rev. Condens. Matter Phys. Phys. {\bf 6}, 15 (2015).

\bibitem{ReviewAbanin}
D.A. Abanin, E. Altman, I. Bloch, M. Serbyn, arXiv:1804.11065.

\bibitem{Imbrie}
J.Z. Imbrie, J. Stat. Phys. {\bf 163}, 998, (2016).

%\bibitem{DeRoeck}
%W. De Roeck and Fran\c{c}ois Huveneers, Phys. Rev. B. {\bf 95}, 155129 (2017).

%\bibitem{Potirniche}
%I.-D. Potirniche, S. Banerjee and E. Altman, arXiv:1805.01465.

%\bibitem{Billy}
%J. Billy, et al., Nature {\bf 453}, 891, (2008).

%\bibitem{Roati}
%G. Roati, et al., Nature {\bf 453}, 895, (2008).

%\bibitem{Sanchez-Palencia}
%L. Sanchez-Palencia and M. Lewenstein, Nature Physics {\bf 6}, 87 (2010). 

\bibitem{Tanzi}
L. Tanzi, et al., Phys. Rev. Lett. {\bf 111}, 115301 (2013).

\bibitem{Schreiber}
M. Schreiber, et al., Science {\bf 349}, 842 (2015). 

\bibitem{Smith}
J. Smith,et al., Nat. Phys. {\bf12}, 907 (2016).

\bibitem{Bordia}
P. Bordia, et al., Phys. Rev. Lett. {\bf 116}, 140401 (2016).

\bibitem{Choi}
J. Choi, et al., Science {\bf 352}, 1547 (2016).

\bibitem{Bordia2}
P. Bordia, et al., Phys. Rev. X {\bf7}, 0141047 (2017).

\bibitem{Rubio}
A. Rubio-Abadal, et al., arXiv:1805.00056.

\bibitem{Nandkishore}
R. Nandkishore, Phys. Rev. B {\bf 90}, 184204 (2014).

\bibitem{Aleiner}
I.L. Aleiner, B.L. Altshuler and G.V. Shlyapnikov, Nat. Phys. {\bf 6}, 900 (2010).

\bibitem{Michal}
V.P. Michal, B.L. Altshuler and G.V. Shlyapnikov, Phys. Rev. Lett. {\bf 113}, 045304 (2014).

\bibitem{Michal2}
V.P. Michal, I.L. Aleiner, B. L. Altshuler and G. V. Shlyapnikov, Proc. Natl. Acad. Sci. U.S.A. {\bf 113}, E4455 -- E4459 (2016).

\bibitem{Gornyi}
I.V. Gornyi, A.D. Mirlin, M. M\"uller and D.G. Polyakov, Annalen der Physik {\bf 529}, 1600365 (2017). 

\bibitem{Bertoli} G. Bertoli, V.P. Michal, B.L. Altshuler and G.V. Shlyapnikov, Phys. Rev. Lett {\bf 121}, 030403 (2018).

\bibitem{walraven} 
O.J. Luiten, M.W. Reynolds, and J.T.M. Walraven, Phys. Rev. A {\bf 53}, 381 (1996).

\bibitem{ketterle} 
W. Ketterle and N.J. Van Druten, Adv. At. Mol. Opt. Phys. {\bf 37}, 181 (1996).

\bibitem{Falco}
G.M. Falco, T. Nattermann and V.L. Pokrovsky, Phys. Rev. B {\bf 80}, 104515 (2009).

\bibitem{Lifshitz}
I.M. Lifshitz, Sov. Phys. Usp. {\bf 7}, 549 (1965).

\bibitem{Zittartz}
J. Zittartz and J.S. Langer, Phys. Rev. {\bf 148}, 741(1966).

%\bibitem{Manai}
%I. Manai, et al.,  Phys. Rev. Lett. {\bf 115}, 240603 (2015) .

\bibitem{Lee}
P.A. Lee and T.V. Ramakrishnan, Rev. Mod. Phys. {\bf 57}, 287 (1985).

\bibitem{Mirlin}
A.D. Mirlin, Phys. Rep., {\bf 326}, 259 (2000).

%\bibitem{gornyi2017spectral}
%Gornyi, I. V., A. D. Mirlin, D. G. Polyakov, and A. L. Burin, Annalen der Physik {\bf 529}, 1600360 (2017).

%%%%%%%%%%%%%%%%
\begin{comment}

\bibitem{chang2016} R. Chang, Q. Bouton, H. Cayla, C. Qu, A. Aspect, C.I. Westbrook, and D. Clement, Phys. Rev. Lett. {\bf 117}, 235303 (2016).

\bibitem{clement} D. Clement, N. Fabbri, L. Fallani, C. Fort, and M. Inguscio, Phys. Rev. Lett. {\bf 102}, 155301 (2009).

\bibitem{ernst} P.T. Ernst, S. Gotze, J.S. Krauser, K. Pyka, D-S. Luhmann, D. Pfannkuche, and K. Sengstock, Nature Physics {\bf 6}, 56 (2010).

\bibitem{Huang}
K. Huang and H.-F. Meng, Phys. Rev. Lett. {\bf 69} 644 (1992).

\bibitem{Meng}
H.-F. Meng, Phys. Rev. B {\bf 49} 1205 (1994).

\bibitem{Nelson}
D.R. Nelson and J.M. Kosterlitz, Phys. Rev. Lett. {\bf 39}, 1201 (1977).

\bibitem{Prokof'ev}
N. Prokof'ev, O. Ruebenacker and B. Svistunov, Phys. Rev. Lett. {\bf 87}, 270402 (2001).

\bibitem{Yukalov}
V. I. Yukalov and R. Graham, Phys. Rev. A {\bf 75}, 023619 (2007).

\bibitem{bktMeasurements}
Z. Hadzibabic et al., Nature {\bf441}, 1118-1121 (2006); P. Kr\"uger, et al., Phys. Rev. Lett. {\bf99}, 040402 (2007); P. Clad\'e, et al., Phys. Rev. Lett. {\bf102}, 170401 (2009).

\bibitem{disorderMeasurements}
M. C. Beeler, M. E. W. Reed, T. Hong, and S. L. Rolston, New J. Phys. {\bf14}, 073024 (2012); B. Allard, et al., Phys. Rev. A {\bf85}, 033602 (2012).

\bibitem{Krinner}
S. Krinner, D. Stadler, J. Meineke, J.-P. Brantut, and T. Esslinger, Phys. Rev. Lett. {\bf110}, 100601 (2013).

\bibitem{boxMeasurements}
A. L. Gaunt, et al., Phys. Rev. Lett. {\bf110}, 200406 (2013); L. Corman, et al, Phys. Rev. Lett. {\bf113}, 135302 (2014); L. Chomaz, et al., Nat. Commun. {\bf6}, 6162 (2015).

\bibitem{Ville}
J. L. Ville, et al., Phys. Rev. A {\bf95} 013632 (2017).

\bibitem{Mathey}
L. Mathey, K.J. G\"unter, J. Dalibard, and A. Polkovnikov, Phys. Rev. A 95 053630 (2017).

\bibitem{BK2014} A. De Luca, B.L. Altshuler, V.E. Kravtsov, and A. Scardicchio, Phys. Rev. Lett. {\bf 113}, 046806 (2014).

\end{comment}

\end{thebibliography}
\end{document}